\documentclass[aps,preprintnumbers,11pt,amsmath,amssymb,nofootinbib]{revtex4}

\usepackage{braket}
\usepackage{float}
\usepackage{graphics,epsfig}
\usepackage{bm}
\usepackage{slashed}
\usepackage{graphicx}
\usepackage{amsmath}
\usepackage{amsfonts}
\usepackage{amssymb}
\usepackage{color}%
\usepackage{dcolumn}
\usepackage[
            pdfstartview=FitH,
            bookmarksnumbered=true,
            bookmarksopen=true,
            colorlinks,
            linkcolor=blue,
            anchorcolor=green,
            citecolor=red
            ]{hyperref}
\usepackage{enumerate}
\usepackage{pdflscape}
\usepackage{tabularx}
\usepackage{subfigure}
\providecommand{\U}[1]{\protect\rule{.1in}{.1in}}

\def\A0{A^{(0)}}

\begin{document}
\baselineskip=0.6 cm \title{Deep learning the holographic black hole with charge}

\author{Jing Tan$^{1,2}$}
\thanks{E-mail address: renxingbug@gmail.com}
\author{Chong-Bin Chen$^{1,2}$}

\affiliation{
$^{1}$Department of Physics, Nanchang University, Nanchang, 330031, China\\
$^{2}$Center for Relativistic Astrophysics and High Energy Physics, Nanchang University, Nanchang 330031, China}

\vspace*{0.2cm}
\begin{abstract}
\baselineskip=0.6 cm
\begin{center}
{\bf Abstract}
\end{center}
We use the deep learning algorithm to learn the Reissner-Nordstr\"{o}m(RN) black hole metric by building a deep neural network. Plenty of data is made in boundary of AdS and we propagate it to the black hole horizon through AdS metric and equation of motion(e.o.m) We label this data according to the values near the horizon, and together with initial data constitute a data set. Then we construct corresponding deep neural network and train it with the data set to obtain the Reissner-Nordstrom(RN) black hole metric. Finally, we discuss the effects of learning rate, batch-size and initialization on the training process.
\end{abstract}

\maketitle
\newpage
\vspace*{0.2cm}

%

\section{Introducton}
The holographic duality\cite{Maldacena:1997re, Gubser:1998bc, Witten:1998qj}, especially the so-called AdS/CFT correspondence, describes the equivalency between a $(d+1)$-dimensional supergravitational system and a d-dimensional gauge quantum field. The holographic duality usually refers to the duality hypothesis proposed by Gerard 't Hooft\cite{'t Hooft:1993} and soon was given a precise string-theory interpretation by Leonard Susskind\cite{Susskind:1994vu}. A particular example of this hypothesis was found by Maldacena\cite{Maldacena:1997re}, which proposed that a strongly coupled $4$-dimensional Yang-Mills gauge theory would be dual to a weakly coupled $5$-dimensional AdS supergravity. This implies that one can use the gravitational theory to deal with strongly coupled systems which perturbation theories cannot manage. On the contrary, we can learn about the information we care about on the gravitational field from the quantum field lived on the boundary of the gravitational system. The idea of the equivalence between gravity and quantum field provides an insight that the spacetime is possibly emerged from boundary theories. In other words, the gravitational system can be constructed if one know the information of the boundary, and figure out what equations it obeys.

On the other hand, deep learning \cite{Hinton, Bengio, LeCun} is a field of machine learning that uses a cascade of multiple layers of nonlinear processing units for feature extraction and transformation and each successive layer uses the output from the previous layer as input \cite{L.Deng}. This algorithm abstracts multiple levels of representation, and forms a hierarchy of data. Hinton proposed an algorithm called back propagation \cite{Rumelhart}, which increases the process of error back propagation on the traditional algorithm. In this way, learning problems such as XOR, which were helpless in the perceptron model\cite{Rosenblatt Frank} in the past, can be solved. In 2006, Hinton adopted unsupervised learning to train the algorithm layer by layer, and used supervised back propagation algorithm for tuning, which solved the problem of vanishing gradient\cite{Hinton}, it makes deep learning algorithm improved and it is basic of our deep neural network. At the same time, with the rapid improvement of computing power and the emergence of big data in recent years, deep leaning architectures such as deep neural network, deep belief network, is rapidly riding. Now they are widely used in signal and information processing, such as speech recognition, computer vision, natural language processing and machine translation.

A natural question raised immediately is wether there are some deep relations between holography and machine leaning. In other words, can one construct the gravity holographically from boundary systems through training the deep neural network? Before people realized this relation, someone have studied the connection between deep learning and the renormalization group of a tensor network\cite{Beny:2013,Mehta:2014}. This is a support that AdS space can be emerged from deep learning because the so-called multiscale entanglement renormalization ansatz(MERA) network was regard as a discrete time slice from holographic point of view\cite{Swingle:2009bg}, and following study motivated by this was discussed in \cite{ Gan:2017nyt}. Recently Hashimoto et.al. have achieved the metric of an AdS black hole via deep neural network with boundary input data\cite{Hashimoto:2018ftp}.  Our work is base on \cite{Hashimoto:2018ftp} and test it into the Reissner-Nordstrom(RN) metric\cite{Reissner:1916, Weyl:1917, Nordstrom:1918, Jeffery:1921}, whose dual boundary theory should contain vector fields. For constructing a neural network the things we should know, as we said, are the input data(boundary value of fields and its derivative) and the equations they obey(equations of motion of these fields). To achieve it, one can obtain related equations of motion(e.o.m) from the gravitational action. After discretizing e.o.m in radial direction and map them into a neural network, the weights and active function will be determined and the network proceed availably after a coordinate transformation $d\eta=dr/\sqrt{f}$. And then we produce data set(see Fig.\ref{fig:data_picture}) and construct specific neural network of scalar field in new coordinate system. After selecting the appropriate regularization term and activation function, we implemented the neural network on TyTorch, which is is an open-source machine learning library for Python and succeed to learn the RN black hole metric through training the network, see Fig.\ref{rn_metric_pic}. In addition, if we consider effects of different parameters on training, we find that the loss will converge to same value as the training goes on, Fig.\ref{rn_loss_pic}. this means that the deficiencies of some parameter setting can be remedied by adding number of epoch.

\section{Metric reproduce of Reissner-Nordstr\"{o}m-AdS black hole by Deep neural Network}


Let us start with the $n$-dimensional Reissner-Nordstr\"{o}m-AdS(RNAdS) black hole
\begin{gather}
ds^2=-f(r)dt^2+\frac{dr^2}{f(r)}+r^2\left(d\varOmega^k_{n-2}\right)^2,
\label{rn_metric} \\
f(r)=k-\frac{8\pi}{(n-2)\varOmega^k_{n-2}}\frac{2GM}{r^{n-3}} + \frac{8\pi}{(n-2)\varOmega^k_{n-2}}\frac{GQ^2}{r^{2(n-3)}} + \frac{r^2}{L_{AdS}^2} \label{rn_f},
\end{gather}
where $L_{AdS}$ is AdS radius, $M$ and $Q$ are the mass and charge of the RNAdS black hole, respectively. The constant $k$ has values $k=0,1,-1$ depending on the topology is planar, spherical, or hyperbolic. The angular part $d\varOmega^k_{n-2}$ of the metric\eqref{rn_metric}, for each $k=0,1,-1$, is of the form
\begin{eqnarray}
\left(d\varOmega^{0}_{n-2}\right)^2&=&\sum_{i=1}^{n-2}d\theta_i^2\\
\left(d\varOmega^{1}_{n-2}\right)^2&=&d\theta_1^2+\sin^2\theta_1d\theta_2^2+\cdots+\prod_{i=1}^{n-3}\sin^2\theta_id\theta_{n-2}^2\\
\left(d\varOmega^{-1}_{n-2}\right)^2&=&d\theta_1^2+\sinh^2\theta_1d\theta_2^2+\cdots+\sinh^2\theta_1\prod_{i=2}^{n-3}\sin^2\theta_id\theta_{n-2}^2
\end{eqnarray}
$\varOmega^k_{n-2}$ denotes the area of the unit surface of each kind.

In this background, we consider a scalar field with Lagrangian density
\begin{equation}
\mathcal{L}=-\frac12(\partial_\mu\phi)^2-\frac12m^2\phi^2-V(\phi),
\end{equation}
which yields the following equation of motion
\begin{equation}
\square\phi-m^2\phi-\frac{\delta V}{\delta \phi}=0.
\end{equation}

Substituting \eqref{rn_metric} we obtain
\begin{equation}
  f\pi' +\left(f'+\frac{2f}{r}\right)\pi-m^2\phi-\frac{\delta V}{\delta \phi}=0,
  \label{rn_eom}
\end{equation}
where $\pi\equiv \phi'(r)$.

Our strategy is to provide a representation of scalar field equation in deep neural network. The emerged holographic direction($\eta$) is discretized and regarded as the deep layers. Naively, to map the e.o.m to deep neural network, we can discretize the equations in radial direction by (\ref{rn_eom})
\begin{eqnarray}
 \phi(r+\Delta r)&=&\phi(r)+\Delta r\pi(r), \nonumber  \\
 \pi(r+\Delta r)&=&\pi(r)-\Delta r \left\{-\left(\frac{f'}{f}+\frac{2}{r}\right)\pi(r)-\frac{1}{f}\left(m^2\phi(r)-\frac{\delta V(\phi)}{\delta\phi(r)}\right)\right\}, \label{rn_discretization_r}
  \end{eqnarray}
Where $\Delta r$ is distance of adjacent points in discrete coordinate system, and $r^{(n)}\equiv(N-n+1)\Delta r$, $N$ is total number of layers of deep neural network. In this black hole model the input data $x_{i}$ in each layer of the neural network are the scalar field $\phi$ and its derivative $\pi$. Between the adjacent layers there are a linear transformation $x_{i}\to W_{ij}x_{j}$ and  a nonlinear transformation $x_{i}\to\varphi(x_{i})$ which is known as the activation function. A conventional deep feedforward neural network can be constructed
\begin{equation}
y(x^{(1)})=f_i \varphi (W_{ij}^{(N)}\varphi (W_{jk}^{(N-1)}\cdots \varphi (W_{lm}^{(1)}x_m^{(1)})\cdots)).
\label{feedforward}
\end{equation}

On $r =\infty $, take the $\phi ~and ~\pi $ in \eqref{rn_discretization_r} as input data $x_m^{(1)}$ put in neural network, and then propagate it to the output data on $r=0$ (black hole horizon) in accordance with equation (\ref{rn_discretization_r}) in each layer. The final layer $f_i$ is for summarizing all the component of $\varphi$ and map to the output data $y$. From (\ref{rn_discretization_r}) it's obvious to find the linear part in neural network is
\begin{equation}
W^{(n)}=\begin{pmatrix}
1&\Delta r\\
\frac{m^2}f &\ \ \ 1-\left(\frac{f'}f+\frac{2}{r}\right)\Delta r
\label{RN_W_r}
\end{pmatrix},
\end{equation}
and the activation function at each layer is given by
\begin{equation}
\begin{cases}
\varphi_1(x_1)=x_1 \\
\varphi_2(x_2)=x_2+\Delta r \frac{\delta V(x_1)}{f\delta \phi}
\label{rn_active_fn_r}
\end{cases}
\end{equation}
Where n is the total number of layers, this process is called forward propagation.

\begin{figure}[H]
  \centering
  \includegraphics[scale=0.3]{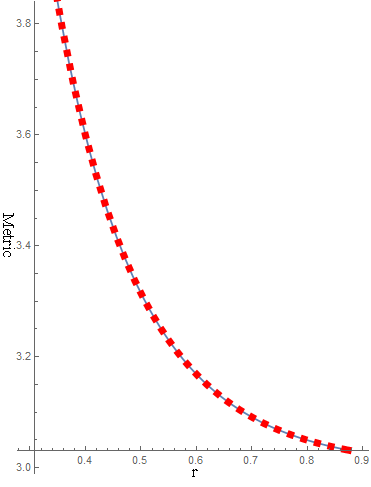}
  \caption{The real line is the numerical result and the dashed line is the analytical one.}
  \label{numerical_integration_picture}
\end{figure}

However, in this coordinate system, we find the metric function $f$ included in the weight function and the activation function if we use \eqref{RN_W_r} and \eqref{rn_active_fn_r} to build a neural network. In that case, we will encounter some technical difficulties that we cannot overcome for the time being. To avoid that, we need a coordinate transformation to make coefficient of radial direction of metric in \eqref{rn_metric} not contain f . This transformation excluded $f$ from coefficient of second-order derivative of scalar field($\partial_r\pi$) in \eqref{rn_eom}, so we obtain a activation function without metric and build an effective neural network. Here is the transformation
\begin{equation}
d \eta= \frac{dr}{\sqrt{f}}.
\label{eta_f}
\end{equation}
After integration we get the coordinate $\eta=G(r)$ and its inverse function $r=G^{-1}(\eta)$. By use this function, we can turn above equations from coordinate $r$ to $\eta$. It turns out in this new coordinate both weight function and activation function are independent of $f$. But for the specific \emph{f} given in $\eqref{rn_f}$, we don't have an analytic expression of integration \eqref{eta_f}. We have to appeal to numeric integrations.  To verify our numerical validity, we compare our numerical results with the ones obtained analytically for Schwarzschild case, as plotted in (fig.\ref{numerical_integration_picture}).

Let us turn to the Reissner-Nordstr\"{o}m-AdS black hole metric reproduced by getting the points from the numerical integration with different charge. Using numerical integration, each value of $r$ reproduces a value of $\eta$ and the corresponding metric. Therefor we construct a metric function of $ \eta $, which called \emph{Reproduced Metric} and is written as $R(\eta)$. The metric constructed in this way is discrete by nature, this saves us much trouble in building neural network. Suppose we get infinite number of points from radius $r$, we will get a continuous reproduced metric, then we can get the e.o.m immediately:
\begin{equation}
     \partial_\eta\pi+R(\eta)\pi-m^2\phi-\frac{\delta V[\phi]}{\delta\phi}=0.\label{eom}
 \end{equation}

Consider finite points case, we can get discrete e.o.m of scalar field:
\begin{gather}
 \phi(\eta+\Delta\eta)=\phi(\eta)+\Delta\eta\pi(\eta), \notag  \\
 \pi(\eta+\Delta\eta)=\pi(\eta)-\Delta\eta(R(\eta)\pi(\eta)-m^2\phi(\eta)-\frac{\delta V(\phi)}{\delta\phi(\eta)}).
 \label{Discretization}
\end{gather}
Where $\Delta \eta$ is distance of adjacent points in discrete coordinate system, and $\eta^{(n)}\equiv(N-n+1)\Delta \eta$, $N$, as before, is total number of layers of deep neural network.
At the same time, we regard the reproduced metric as the target that our neural network learn, and use \eqref{feedforward} as the neural network structure.

Mapping the discrete e.o.m into deep neural network we have weights
 \begin{equation}
 W_{(n)}=\begin{pmatrix}
1&\Delta\eta\\
\Delta\eta m^2 &1-\Delta\eta R(\eta^{(n)})
\end{pmatrix}
\end{equation}
and activation functions in each layer
 \begin{equation}
 \begin{cases}
 \varphi_1(x_1)=x_1; \\
 \varphi_2(x_2)=x_2+\Delta \eta\frac{\delta V(x_1)}{\delta x_1}.
 \end{cases}
 \end{equation}

\begin{figure}[!htbp]
\centering
\includegraphics[scale=0.5]{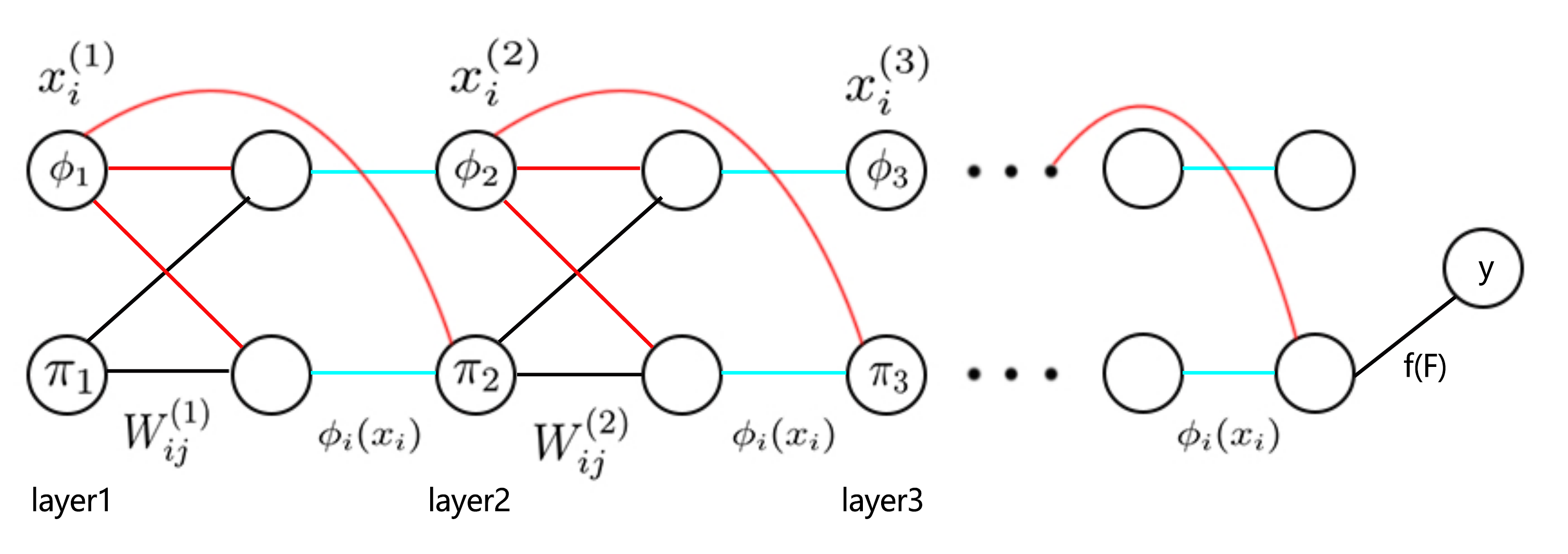}
\caption{Structure of neural network.}
\label{net_str}
\end{figure}

With these elements, we have the structure of the entire network (see Fig.\ref{net_str}). The output of network will generate a loss function E. The information of the loss function will be propagated back by back propagation algorithm to calculate the gradient. Replicating this process layer by layer we obtain the emergent black hole metric.

\section{Methods}

In the last section we have built a deep neural network model of Reissner-Nordstr\"{o}m-AdS black hole. In this section we show more details to implement the network. For simplicity, consider 3+1 dimension spacetime in the unit $L=1$, $m^2=-1$ and $V[\phi]=\frac{\lambda\phi^4}{4}$(where$\lambda=1$)\footnote{Actually, more generic $\phi^n$($n>2$ is an integer) has the similar results.}case, and set $N=10$ i.e. $10$ layers neural network, then $\eta_{ini}=1, \eta_{fin}=0.1$ and $\Delta\eta=-0.1$.
\subsection{Generating the data set}

\begin{figure}[H]
  \centering
  \includegraphics[scale=0.4]{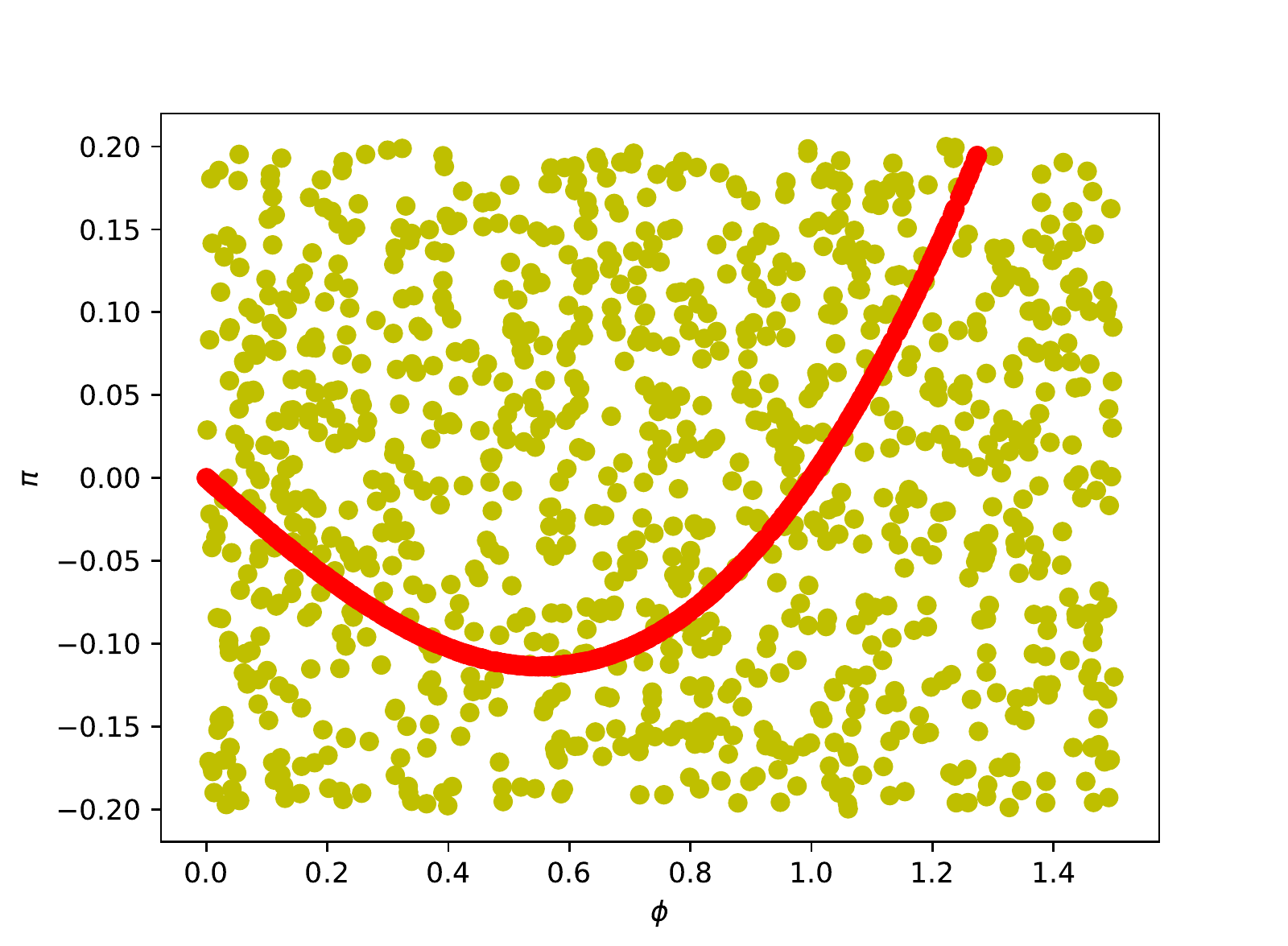}
  \caption{The ($\pi -\phi$) graph of data. The yellow points correspond to the negative data($y=1$) and the red ones are the positive data($y=0$).}
  \label{fig:data_picture}
\end{figure}

Training data is an indispensable part of deep learning. 
Following \cite{Hashimoto:2018ftp}, our data are generated by propagating the e.o.m for a given metric. These data are treated as the data set of the neural network, which can be used to training the network as will see below.  More specifically, we first get data randomly from ($\phi\in[0, 1.5], \pi\in[-0.2, 0.2]$) in the location of $\eta=1$ (i.e. AdS boundary $\eta_{ini}$) as the initial response data. We then propagate data from $\eta_{ini}$ to $\eta_{fin}$(i.e. black hole horizon) by using the e.o.m\eqref{eom}. As the last step, we use the boundary condition of the black hole horizon
\begin{equation}
0=F\equiv\left[\frac{2}{\eta}\pi-m^2\phi-\frac{\delta V(\phi)}{\delta\phi}\right]_{\eta=\eta_{fin}}
\label{boundary_condition}
\end{equation}
 to divide the data into two categories labeled by 0 and 1. The positive answer data characterized by  \newcommand*\abs[1]{\lvert#1\rvert} $\abs{F}<0.1$ are labeled by 0, and the negative answer data characterized by $\abs{F}>0.1$ is labeled by 1. We take the combination of initial data $(\phi(\eta(ini)), \pi(\eta(ini)))$ and corresponding label as a data point. Taking 1000 positive and 1000 negative data points forms the data set. This 2000 data satisfy the distribution as shown in Fig.\ref{fig:data_picture}, which is comparable with the results of the Schwarzschild case in \cite{Hashimoto:2018ftp}.

\subsection{Training the Neural Network}

\begin{figure}[H]
  \centering
  \includegraphics[scale=0.4]{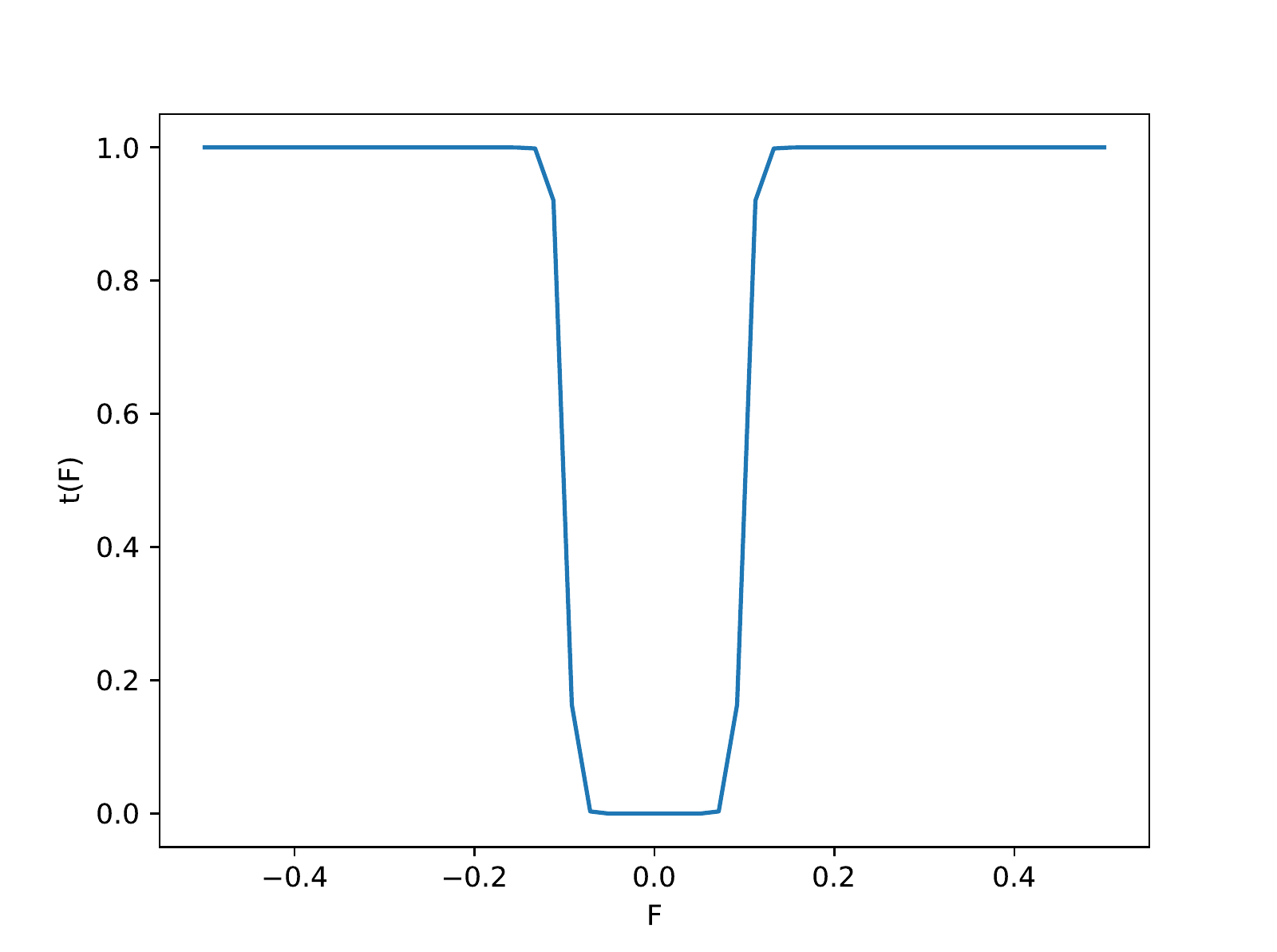}
  \caption{Activation function $f(F)$.}
  \label{active_function_pic}
\end{figure}

We take the data obtained in the last subsection as a training set. The metric is expected to be obtained by training the network. Our training tool is PyTorch \cite{Ketkar,Paszke}, which is a powerful open source machine learning library of Python. We use it to construct a 10 layers deep neural network to train the data. The training data are imported into the deep neural network after divided into 200 batches. Then the neural network transports the data from input layer to the final ($\eta_{fin}$). Following \cite{Hashimoto:2018ftp}, choosing
\begin{equation}
f(F)=[tanh(100(F-0.1)-tanh(100(F+0.1))+2]/2
\label{active_function}
\end{equation}
as the activation function is effective. Here $F$ is one of final layer neuron(in the limit $\eta_{fin}\to0$, equation \eqref{boundary_condition} is equivalent to F$\equiv\pi(\eta_{fin})$). This function $y=f(F)$ is visualized in FIG.\ref{active_function_pic}.

So far, we construct a deep feedforward neural network describe by \eqref{feedforward}. The neural network transmits the data through whole network and generates the output data, which is the actual value. As a contrast, the label is called the target value. The difference between actual value and target value is the loss, which is described by the loss function\cite{Hashimoto:2018ftp}
\begin{equation}
E=E_1+E_{reg}=\sum_{data}|y(\overline x^{(1)})-\overline y| +E_{reg}.
\label{loss_function_all}
\end{equation}

Where $E_1 ~is~ L^1-norm$ , $E_{reg}$ is regularizer and ${( \overline x^{(1)}, \overline y)}$ are the training set. The smaller loss means the closer proximity between the value given by neural network and target value.

\section{Display of Parameters and Training Result }
\begin{figure}[!htbp]
\centering
\subfigure[1]{
\label{rn_fig1} 
\includegraphics[width=7cm,height=4.5cm]{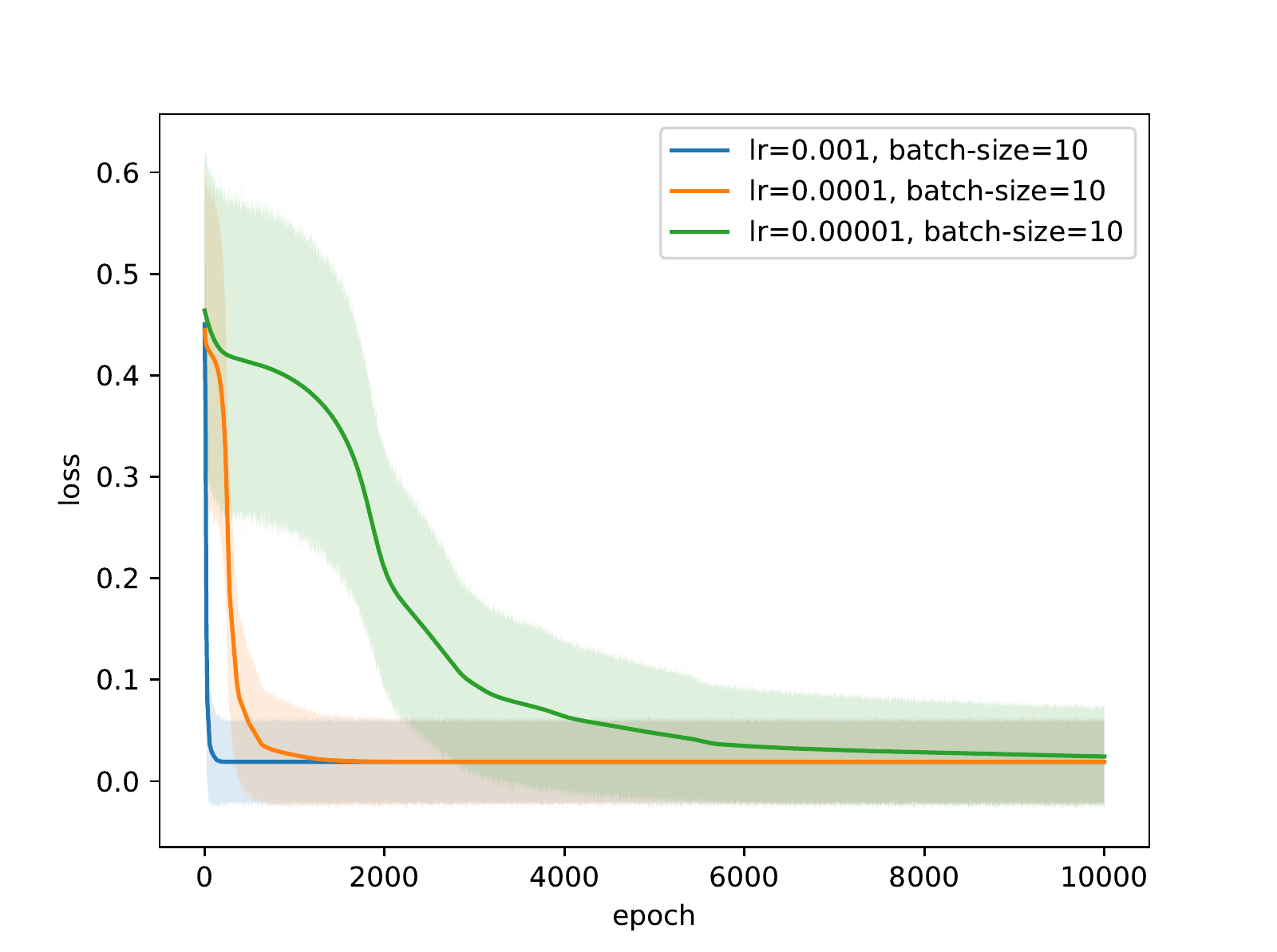}}
\hspace{0.5in}
\subfigure[2]{
\label{rn_fig2} 
\includegraphics[width=7cm,height=4.5cm]{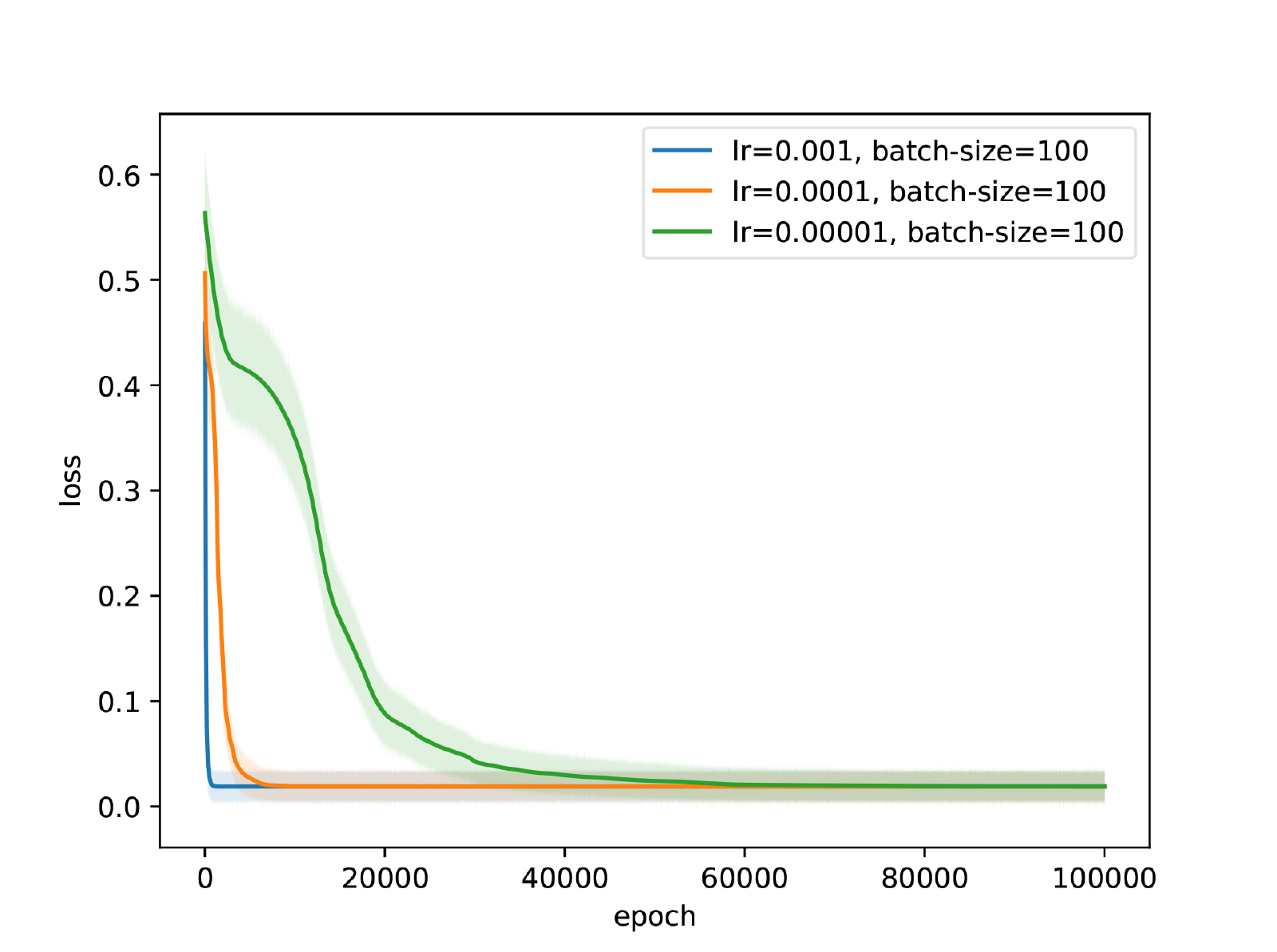}}
\hspace{0.5in}
\subfigure[3]{
\label{rn_fig3} 
\includegraphics[width=7cm,height=4.5cm]{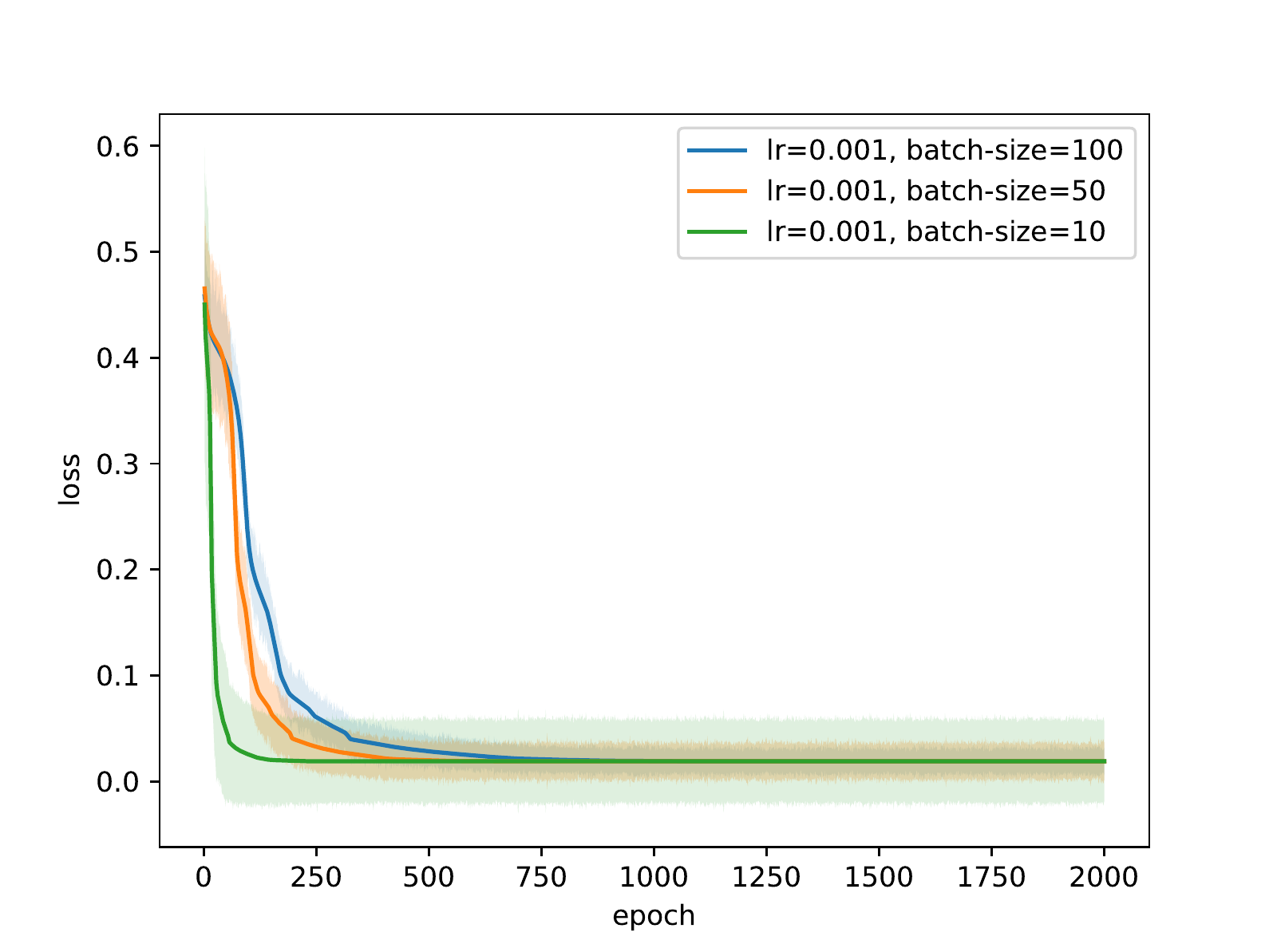}}
\hspace{0.5in}
\subfigure[4]{
\label{rn_fig4} 
\includegraphics[width=7cm,height=4.5cm]{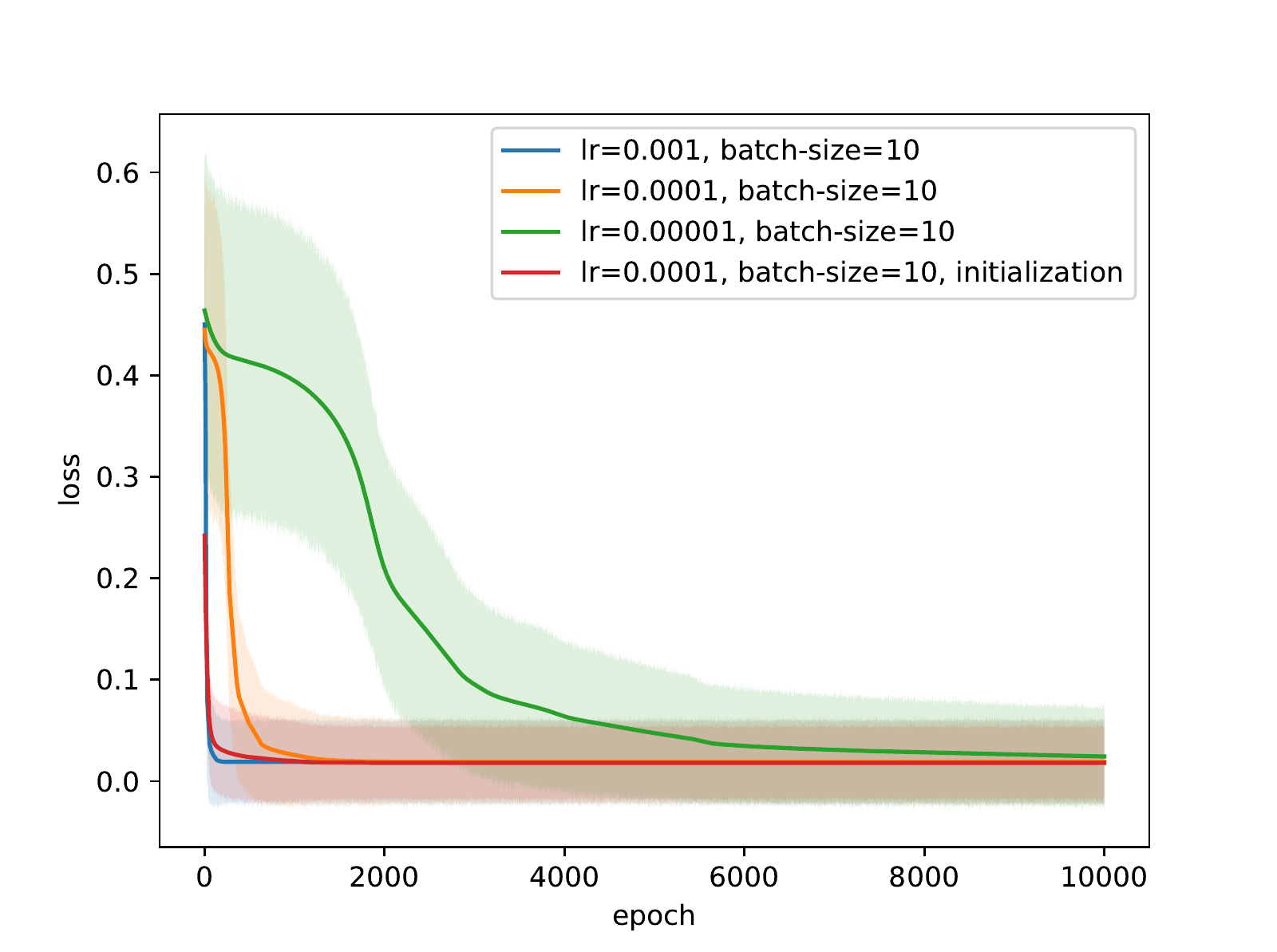}}
\caption{The influence of learning rate and batch size on loss in RN case.}
\label{rn_loss_pic} 
\end{figure}

\begin{figure}[!htbp]
\centering
\subfigure[1]{
\label{rn_fig1} 
\includegraphics[width=7cm,height=4.5cm]{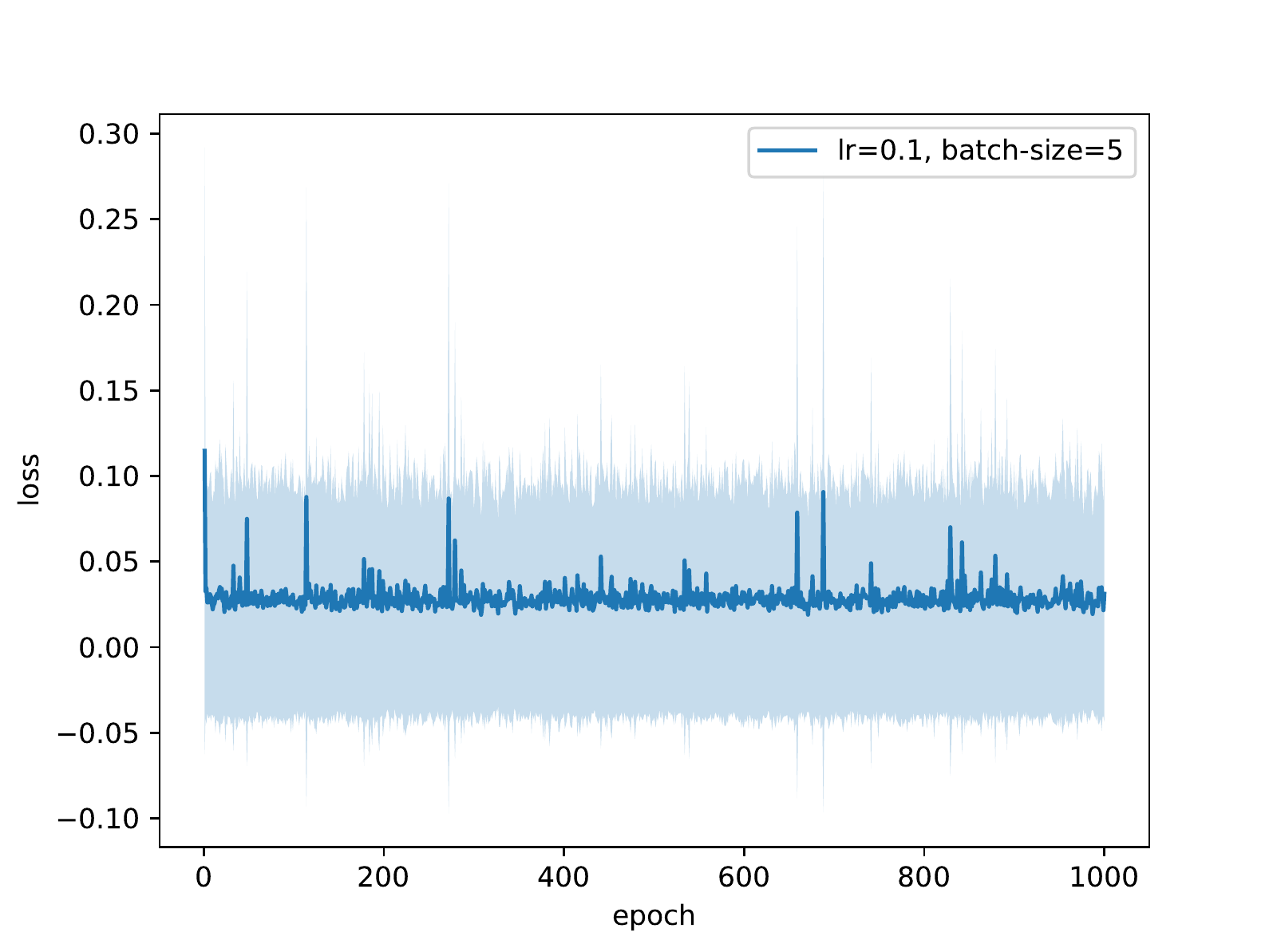}}
\hspace{0.5in}
\subfigure[2]{
\label{rn_fig2} 
\includegraphics[width=7cm,height=4.5cm]{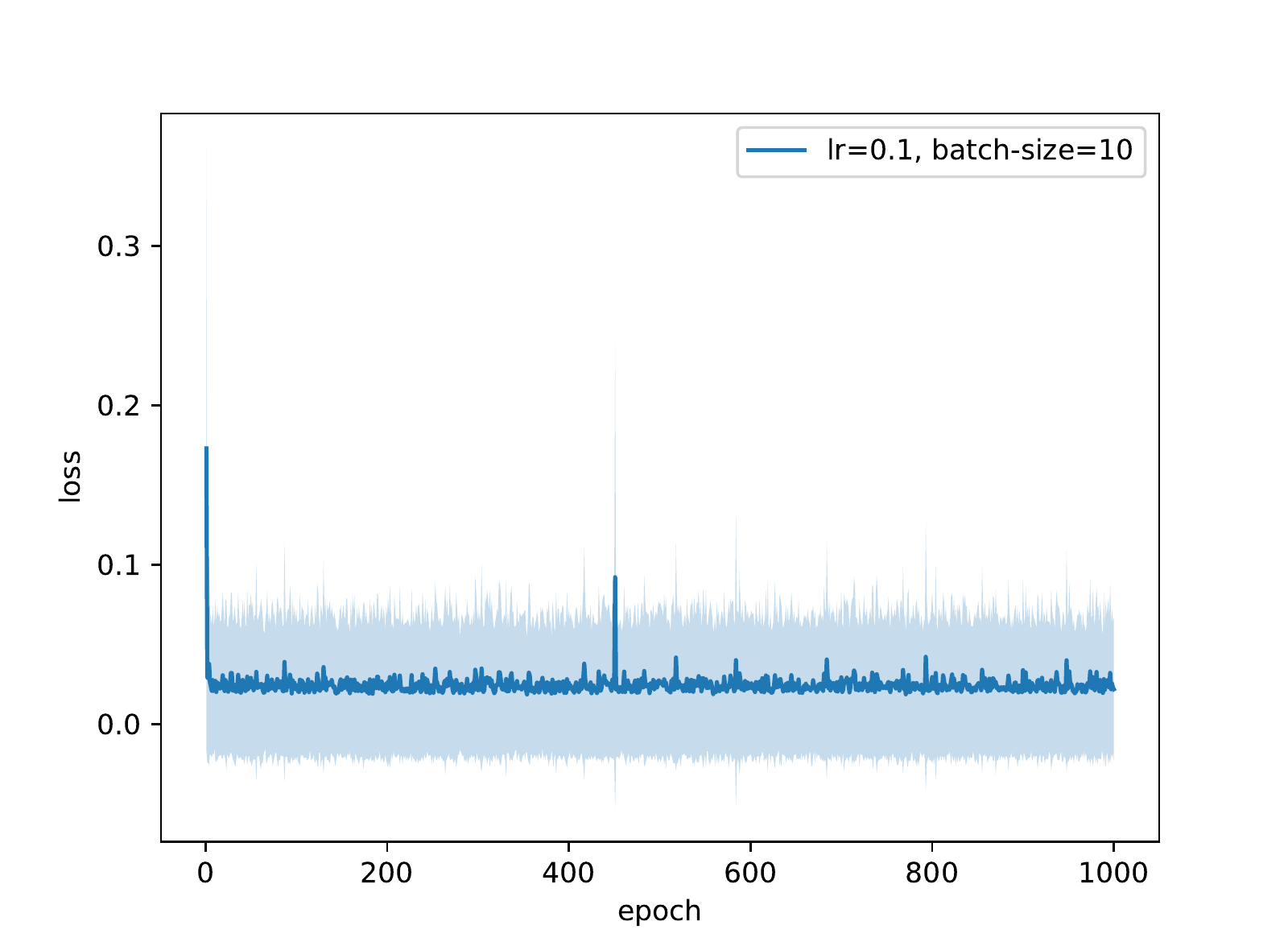}}
\hspace{0.5in}
\caption{The loss oscillation lead by high learning rate.}
\label{rn_lr} 
\end{figure}

In order to reduce the loss, we optimize the neural network weights by Adam optimizer, that is a back propagation algorithm. Optimizer contains a parameter which controls convergence rate of the loss, which is called learning rate or stride. Generally, small learning rate leads to slow convergence.  However, Excessive learning rate has a negative impact on convergence as well. In that case loss may over the bottom and get a suboptimal result. Additionally, choosing suitable batch size is also very important to give an ideal accuracy of the network during training. A larger batch size makes better use of the GPU parallelism, but compare to a smaller one, it require training the network for more epochs to achieve the same level of accuracy. You can set batch size to 1 if you can tolerate the training time. Their impacts are plotted in FIG.\ref{rn_loss_pic}. From these figures, we notice that by choosing appropriate learning rate and batch size, loss will converge to same value as long as the number of epoches is sufficient. And we find that too high value of learning rate in our neural network will make the loss cannot be convergent, like FIG.\ref{rn_lr}. It turns out that the optimal learning rate is around 0.001.

\begin{figure}[!htbp]

\centering
\subfigure[RN1]{
\label{fig:subfig:a} 
\includegraphics[width=3.5cm,height=4cm]{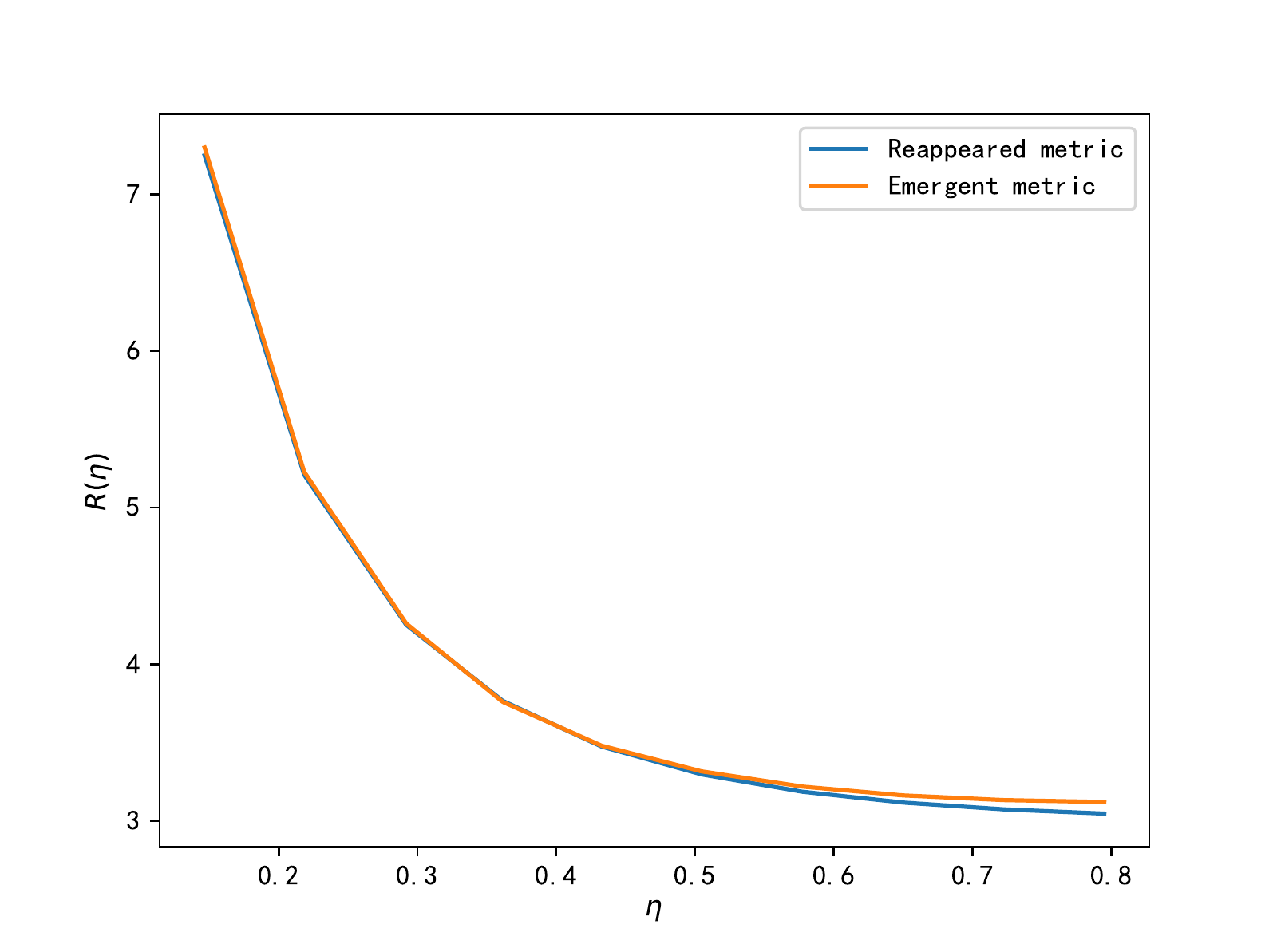}}
\hspace{0.5in}
\subfigure[RN2]{
\label{fig:subfig:b} 
\includegraphics[width=3.5cm,height=4cm]{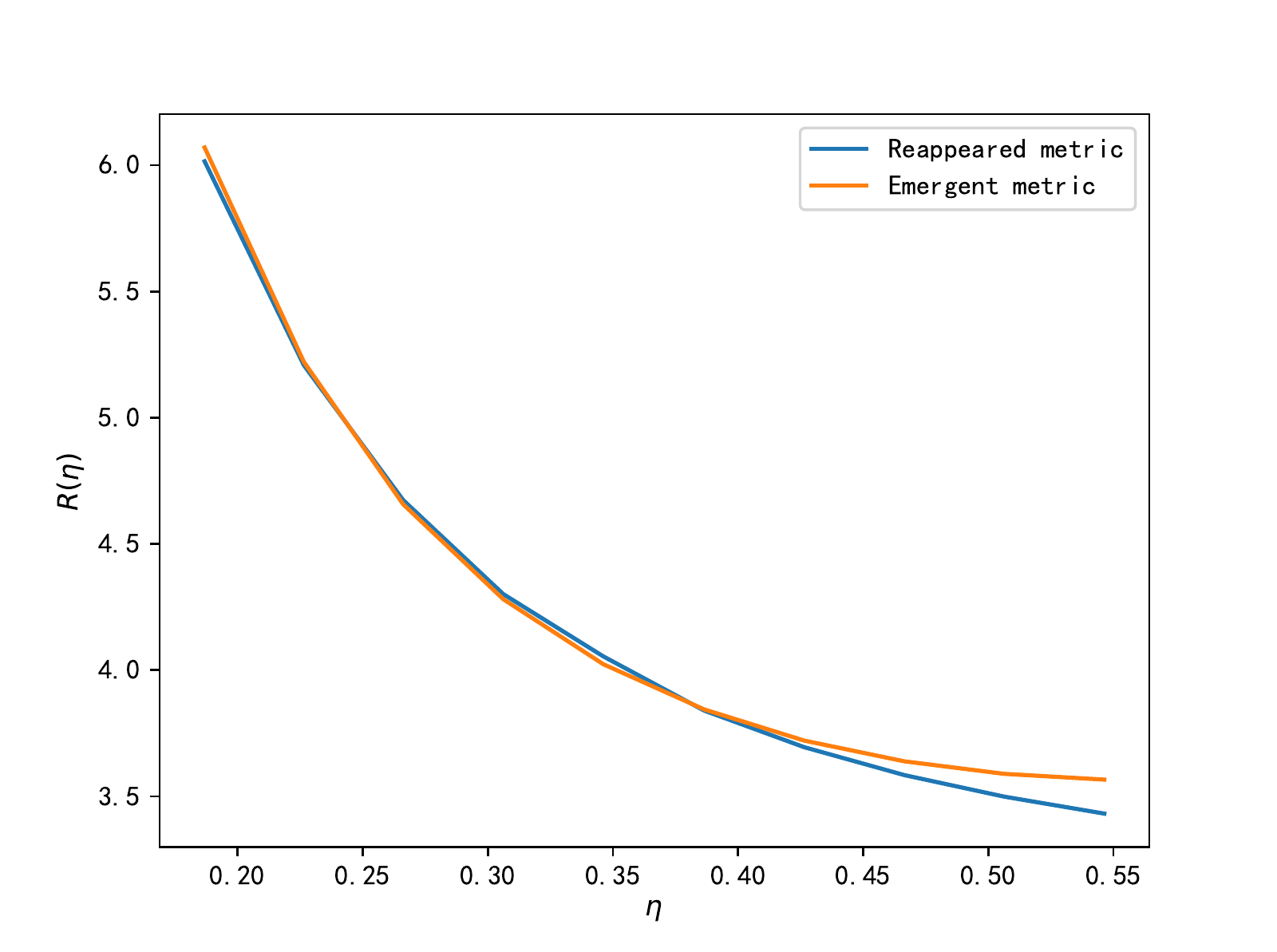}}
\hspace{0.5in}
\subfigure[RN3]{
\label{fig:subfig:c} 
\includegraphics[width=3.5cm,height=4cm]{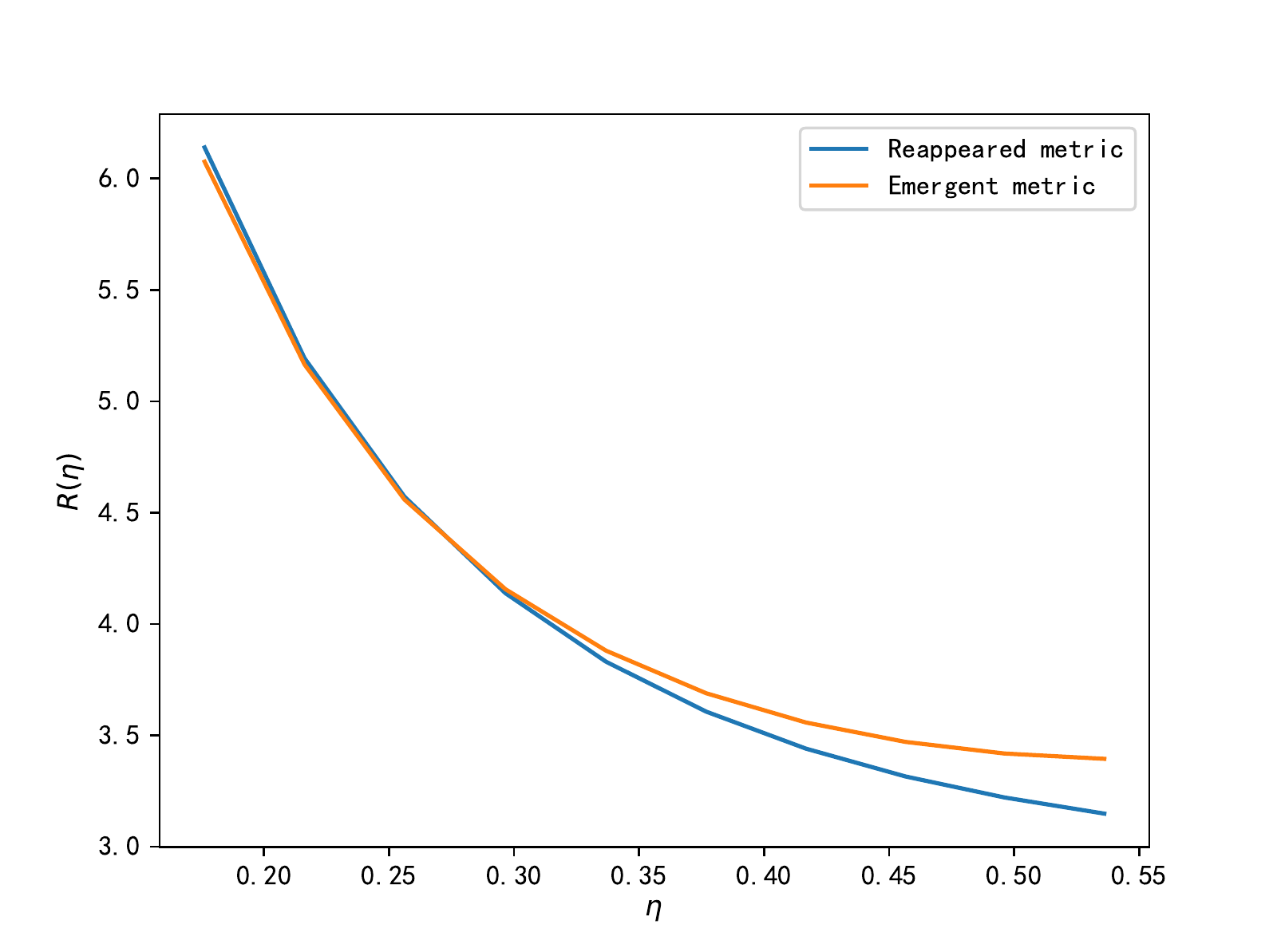}}
\hspace{0.5in}
\subfigure[RN4]{
\label{fig:subfig:c} 
\includegraphics[width=3.5cm,height=4cm]{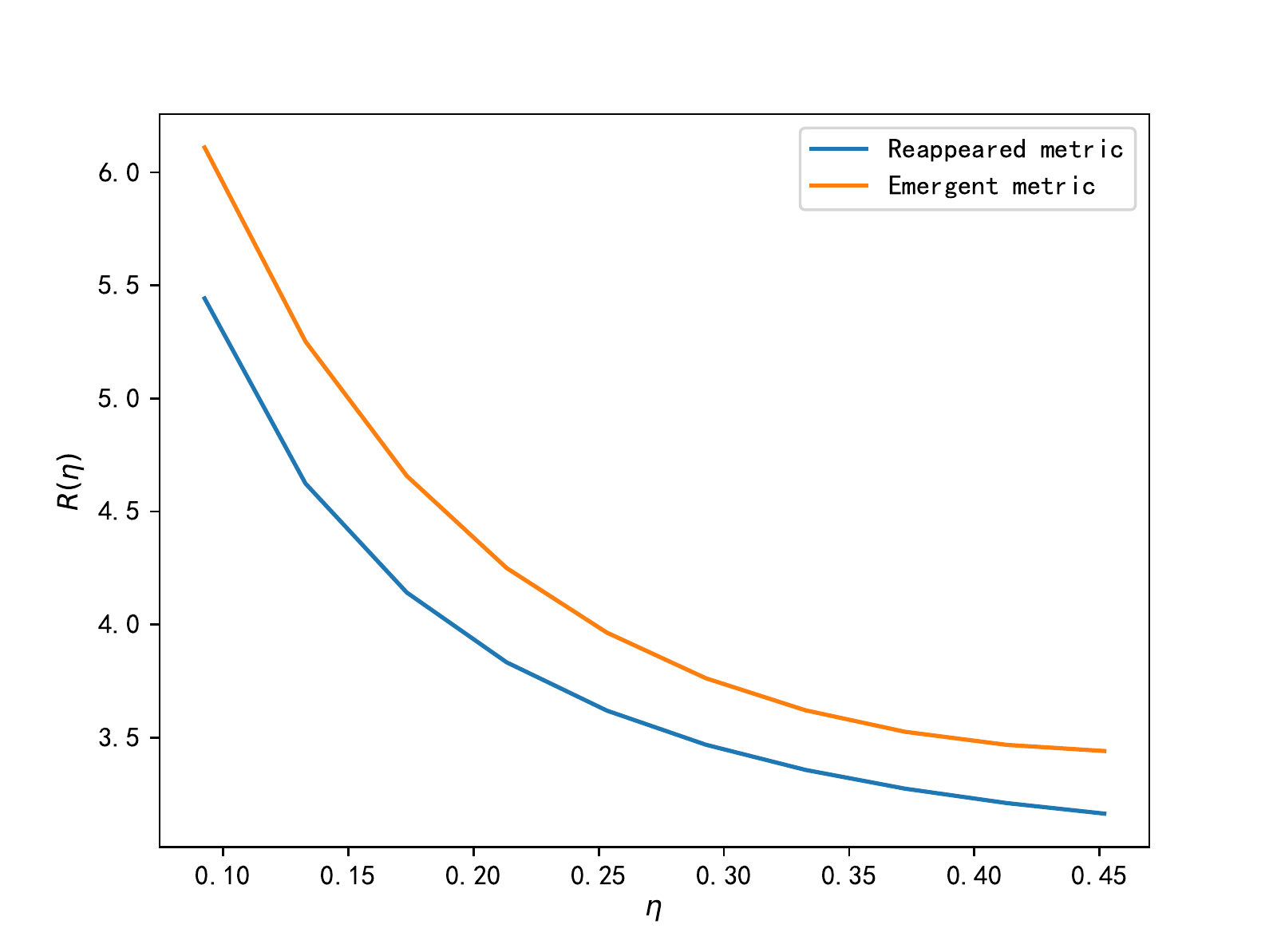}}
\hspace{0.5in}
\subfigure[RN5]{
\label{fig:subfig:c} 
\includegraphics[width=3.5cm,height=4cm]{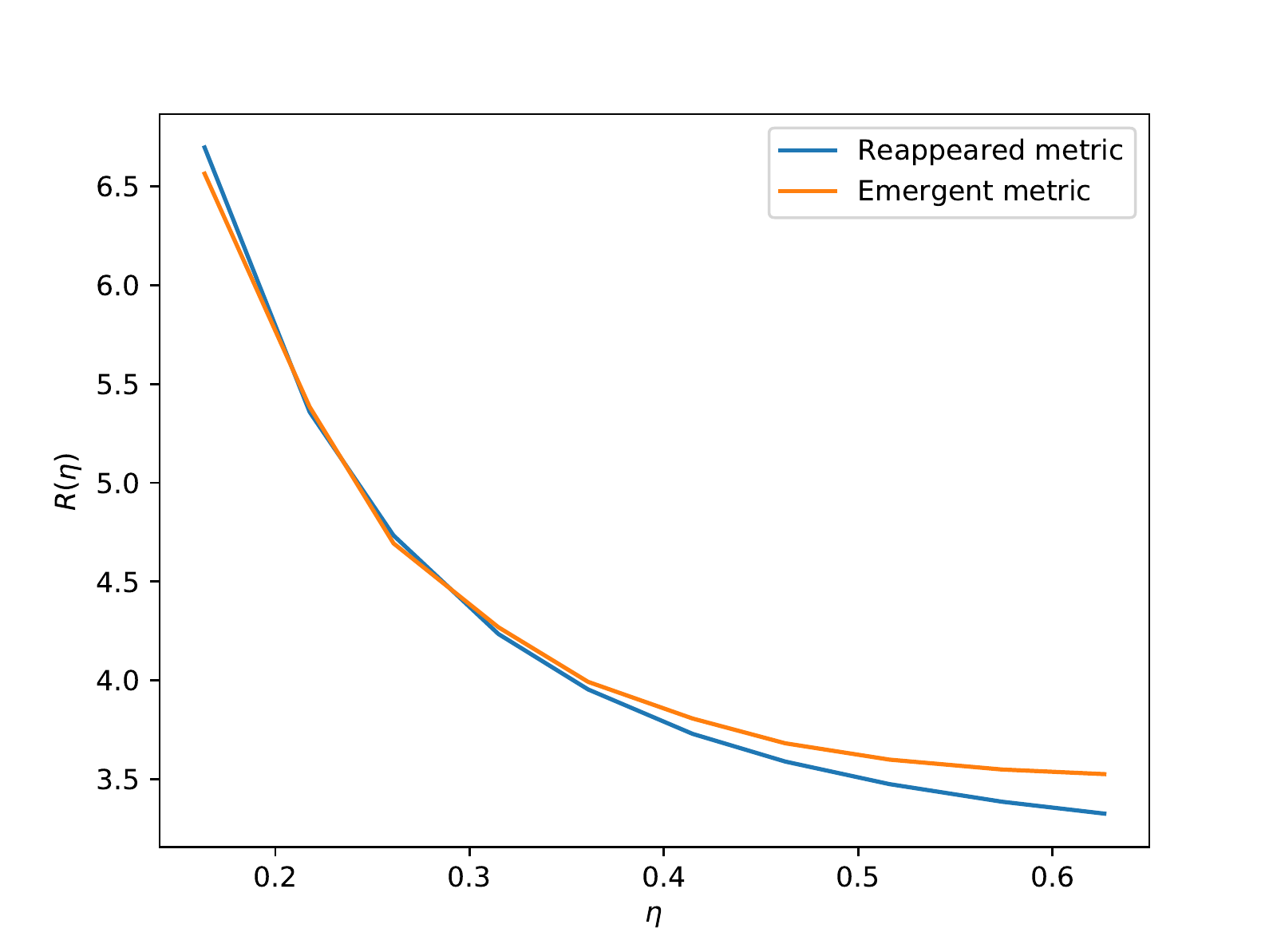}}
\hspace{0.5in}
\subfigure[RN6]{
\label{fig:subfig:c} 
\includegraphics[width=3.5cm,height=4cm]{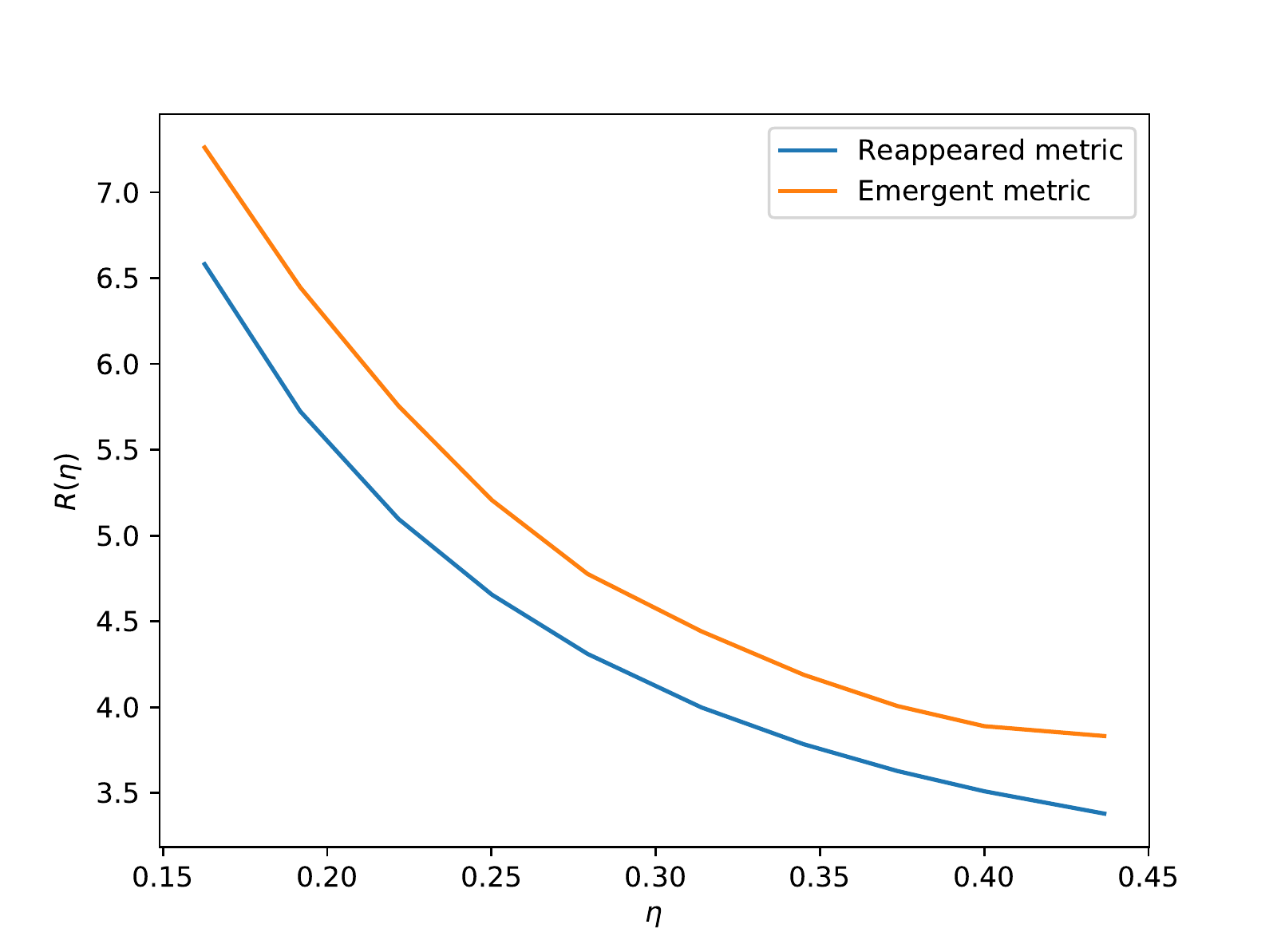}}
\caption{Emerge RN-AdS metric for different charge $Q$ and topology $k$. Details of each case RN1$\cdots$ RN5 are listed in Table I. }
\label{fig:subfig} 

\label{rn_metric_pic}
\end{figure}

Consider the training time and effect, we set learning rate to 0.001 and make batch size equal to 10 to train the neural network. According to experience, we find the appropriate regularization term values of different black holes. In FIG.\ref{rn_metric_pic} we see that RN metric can be learned by training our neural network for any given charge $Q$ and topology $k$ as shown in TABLE I. In order to demonstrate the results of neural network better, in  appendix material, we give out a comparison between the discrete metric which is from the neural network and the standard one, we also give out the mean square error and its change in the program progresses.

\begin{table}[!htbp]
\caption{Parameters of each RN black holes.}
\begin{tabularx}{\textwidth}{XXXc}
\hline
Black holes&Q&k&Regularizer \\ \hline\hline
RN1&0.5&0& $0.039\sum_{n=1}^{N-1}(\eta^{(n)})^{3.6}(h(\eta^{(n+1)})-h(\eta{(n)}))^2$ \\ \hline
RN2&0.5&1& $0.05\sum_{n=1}^{N-1}(\eta^{(n)})^{2.6}(h(\eta^{(n+1)})-h(\eta{(n)}))^2$ \\ \hline
RN3&0.5&-1& $0.0225\sum_{n=1}^{N-1}(\eta^{(n)})^{2.4}(h(\eta^{(n+1)})-h(\eta{(n)}))^2$ \\ \hline
RN4&0.25&0& $0.05\sum_{n=1}^{N-1}(\eta^{(n)})^{2.5}(h(\eta^{(n+1)})-h(\eta{(n)}))^2$ \\ \hline
RN5&0.25&1& $0.028\sum_{n=1}^{N-1}(\eta^{(n)})^3(h(\eta^{(n+1)})-h(\eta{(n)}))^2$ \\ \hline
RN6&0.25&-1& $0.0115\sum_{n=1}^{N-1}(\eta^{(n)})^2(h(\eta^{(n+1)})-h(\eta{(n)}))^2$ \\
\hline
\end{tabularx}
\end{table}

\section{conclusion}

In this paper we apply the deep learning algorithm to black hole holography. We build a neural network to learn the RN-AdS metric of different topology. It turns out the expected metric can be obtained by training this network. It implies, according to AdS/CFT correspondence, that by learning the CFT boundary data, deep neural network reproduces the AdS metric even for spacetime with charge and different topology. This shed new insight into the possible relation between deep learning and the AdS/CFT correspondence. Technically, compared with the Schwarzschild case, RN black hole has some subtleties when build the neural network in terms of the equations of motion. We develop a numeric method to overcome this difficulty. As a last point, we also discuss the influence of different learning rate, batch size and epoch to the neural network. We find there is an optimal choice of these parameters.

\section*{\bf Acknowledgements}

The authors would like to acknowledge Ming-Yuan Luo and Tian-Wei Yan for helping construct early dataset and neural network. The author is also grateful to Prof. Fu-Wen Shu for the useful discussions and modification. This work was supported in part by Natural Science Foundation of China under Grants No.11465012 and 11665016,and 555 talent project of Jiangxi Province.

\newpage

\appendix
\section{Comparison and MSE}

\begin{figure}[H]
\centering
\includegraphics[scale=1]{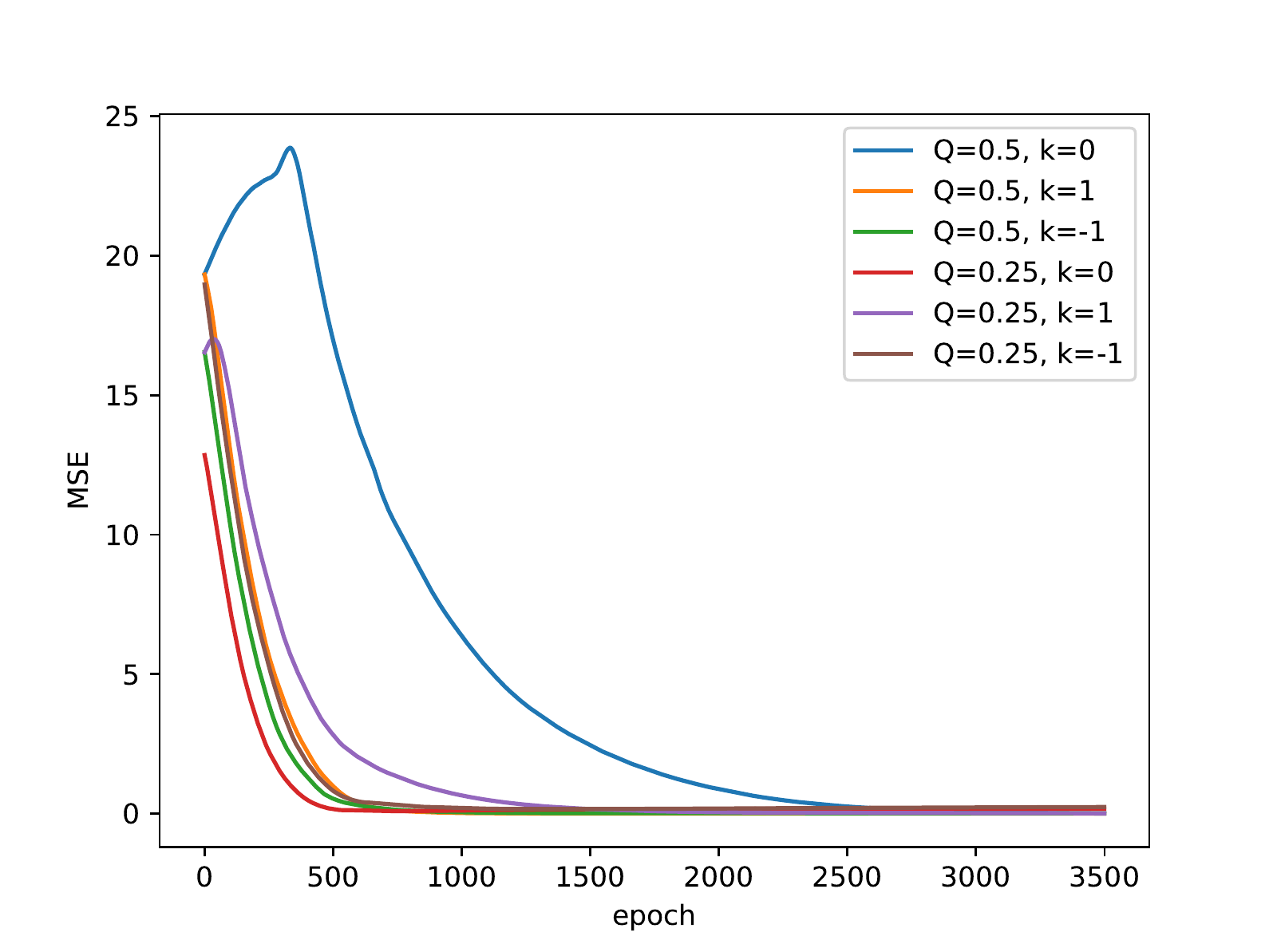}
\caption{The change of mean square error when program running.}
\label{mse_ch}
\end{figure}

As a direct comparison of result, TABEL \ref{comp} contrasts between the discrete metric which is from the neural network and the standard case. In the table, R is the reproduce metric, the em is the emergent metric, and the MSE is the mean square error. In more details, we plot the evolution of the MSE, see Fig.\ref{mse_ch}.

\begin{table}[!htbp]
\resizebox{\textwidth}{16mm}{
\begin{tabular}{c|c|ccccccccccc}
\toprule
&
r &  1.78& 1.68& 1.59& 1.51& 1.44& 1.38& 1.33& 1.29& 1.257& 1.234 \\
RN1&R &3.04501& 3.07225& 3.11558& 3.18464& 3.29514& 3.47341& 3.76487& 4.25044& 5.20848& 7.249\\
&em &3.1193&  3.1314&  3.1612&  3.2175&  3.3147&  3.4787&  3.7586&  4.2587&  5.2273&  7.2988\\
&MSE&&&&&0.00155&&&&& \\
\botrule
\end{tabular}}
\end{table}

\begin{table}[!htbp]
\resizebox{\textwidth}{16mm}{
\begin{tabular}{c|c|ccccccccccc}
\toprule
&
r &  1.192& 1.155 & 1.121 & 1.0896& 1.061  &1.035&  1.012& 0.9918&  0.9745& 0.96 \\
RN2&R &3.43207&3.49943&3.58454&  3.69503&  3.84011&4.03586&4.30142& 4.67335& 5.20802& 6.01493 \\
&em &3.5666&  3.5894&  3.6390&  3.7211&  3.8447&  4.0238&  4.2814&  4.6571&  5.2214&  6.0699\\
&MSE&&&&&0.00347&&&&& \\
\botrule
\end{tabular}}
\end{table}

\begin{table}[!htbp]
\resizebox{\textwidth}{16mm}{
\begin{tabular}{c|c|ccccccccccc}
\toprule
&
r &  1.7645& 1.725& 1.689& 1.656& 1.626& 1.599& 1.575& 1.554& 1.5363& 1.5216 \\
RN3&R &3.14846& 3.22101& 3.31542& 3.44016& 3.60633& 3.83039& 4.13805& 4.57167& 5.19288& 6.13992 \\
&em &3.3945&  3.4184&  3.4706&  3.5575&  3.6889&  3.8801&  4.1559&  4.5589&  5.1649&  6.0762\\
&MSE&&&&&0.01520&&&&& \\
\botrule
\end{tabular}}
\end{table}

\begin{table}[!htbp]
\resizebox{\textwidth}{16mm}{
\begin{tabular}{c|c|ccccccccccc}
\toprule
&
r &  1.589& 1.545& 1.502&  1.463& 1.427& 1.394& 1.3635& 1.336& 1.311& 1.289\\
RN4&R &3.16346& 3.20997& 3.27395& 3.35703& 3.46822& 3.61924& 3.8329& 4.14157& 4.62383& 5.44262 \\
&em &3.4405&  3.4675&  3.5256&  3.6208&  3.7623&  3.9644&  4.2501&  4.6576&  5.2521&  6.1104\\
&MSE&&&&&0.17626&&&&& \\
\botrule
\end{tabular}}
\end{table}

\begin{table}[!htbp]
\resizebox{\textwidth}{16mm}{
\begin{tabular}{c|c|ccccccccccc}
\toprule
&
r &  1.36& 1.3& 1.24& 1.19& 1.15& 1.11& 1.08& 1.05& 1.03& 1.01 \\
RN5&R &3.32611& 3.38634& 3.47534& 3.58949& 3.72965& 3.95443& 4.23485& 4.7336& 5.35848& 6.69649 \\
&em &3.5260&  3.5493&  3.5993&  3.6819&  3.8073&  3.9928&  4.2690&  4.6936&  5.3843&  6.5640\\
&MSE&&&&&0.01189&&&&& \\
\botrule
\end{tabular}}
\end{table}

\begin{table}[!htbp]
\resizebox{\textwidth}{16mm}{
\begin{tabular}{c|c|ccccccccccc}
\toprule
&
r &  1.7& 1.67& 1.65& 1.63& 1.61& 1.59& 1.575& 1.562& 1.55& 1.54   \\
RN6&R &3.38016& 3.51052& 3.62847& 3.78439& 3.99875& 4.31046& 4.65601& 5.09654& 5.72376& 6.58145  \\
&em &3.8322&  3.8897&  4.0071&  4.1890&  4.4421&  4.7764&  5.2068&  5.7551&  6.4465&  7.2612\\
&MSE&&&&&0.27903&&&&& \\
\botrule
\end{tabular}}
\caption{Result display and MSE}
\label{comp}
\end{table}

\end{document}